\documentclass[twocolumn,showpacs,floatfix,prl,superscriptaddress]{revtex4}

\usepackage{float}
\usepackage{graphicx}

\newcommand{\da}{{\dagger}}
\newcommand{\tk}{{\tilde k}}
\newcommand{\beq}{\begin{eqnarray}}
\newcommand{\eeq}{\end{eqnarray}}
\newcommand{\beqq}{\begin{eqnarray*}}
\newcommand{\eeqq}{\end{eqnarray*}}

\begin{document}

\title{Bulk Entanglement Spectrum Reveals Quantum Criticality within a Topological State}

\author{Timothy H. Hsieh}  
\affiliation{Department of Physics, Massachusetts Institute of Technology, Cambridge, MA 02139}
\author{Liang Fu}
\affiliation{Department of Physics, Massachusetts Institute of Technology, Cambridge, MA 02139}


\begin{abstract}
A quantum phase transition is usually achieved by tuning physical parameters in a Hamiltonian at zero temperature.  
Here, we demonstrate that the ground state of a topological phase itself encodes critical properties of its transition to a trivial phase. 
To extract this information, we introduce a partition of the system into two subsystems both of which extend throughout the bulk in all directions. 
The resulting {\it bulk} entanglement spectrum has a low-lying part that resembles the excitation spectrum of a bulk Hamiltonian, which allows us to probe a topological phase transition from a single wavefunction by tuning either the geometry of the partition or the entanglement temperature.  As an example, this remarkable correspondence between topological phase transition and 
entanglement criticality is rigorously established for integer quantum Hall states.  
\end{abstract}

\maketitle

Topological phases of matter are characterized by quantized physical properties that arise from topological quantum numbers. For instance, the quantized Hall conductance of an integer quantum Hall state is determined by its Chern number\cite{tknn}, the quantized magnetoelectric response of a topological insulator is governed by its $Z_2$ topological invariant\cite{hasankane, qizhang, moore}, and the quasi-particle charge in a fractional quantum Hall state is deeply related to its topological degeneracy\cite{niuwen}. 
Remarkably, the complete set of topological quantum numbers, or topological order\cite{wen},  is entirely encoded in the ground state wavefunction, which  
can be computed either directly\cite{tknn, kanemele} or from topological entanglement entropy\cite{levinwen, kitaev, hamma1, hamma2, zhang, spin}.  

To extract more information about a topological phase from its ground state wavefunction, Li and Haldane \cite{lihaldane} considered the full entanglement spectrum of the reduced density matrix upon tracing out a subsystem.  When a topologically nontrivial  ground state is spatially divided into two halves, the resulting entanglement spectrum bears a remarkable similarity to the edge state spectrum of the system in the presence of a physical boundary\cite{bernevig, qi, swingle, pollman, fid, yao, dubail}.   
Given this capability of the entanglement spectrum to simulate {\it edge} excitations,  one may wonder if universal {\it bulk} properties of  topological phases can also be obtained via entanglement.

In this work, we demonstrate that the ground state of a topological phase (a single wavefunction)
encodes information on its phase transition to a trivial product state, despite the fact that the system itself is away from the phase transition.  
To expose this  ``hidden'' topological phase transition, we introduce a new type of real-space partitions which divide the system into two parts that 
are {\it extensive} with system size in all directions, as shown in Fig.1. 
The entanglement Hamiltonian obtained from such an extensive partition is a bulk entity, which we term bulk entanglement Hamiltonian. 
The corresponding bulk entanglement spectrum (BES) has a low-lying part that resembles the excitation spectrum of a physical system in the bulk.  
This enables us to probe bulk properties of topological phases and topological phase transitions,  which cannot be accessed from the left-right partition.  

Our results can be understood from the Chalker-Coddington network model\cite{cc}, which describes the transition between integer quantum Hall states in terms of the percolation of chiral edge states. Our extensive partition generates a periodic array of boundaries. In the case of the $\nu=1$ 
quantum Hall state, each boundary introduces a corresponding edge mode in the bulk entanglement spectrum. 
By varying the geometry of the partition as shown in Fig.1, these edge modes interconnect and percolate throughout the entire system, which {\it mirrors} edge state percolation in an actual quantum Hall transition. This correspondence explains why universal critical properties of a topological phase transition are encoded in a single wavefunction away from criticality, and how they can be extracted from the bulk entanglement spectrum.

\begin{figure*}
\centering
\includegraphics[height=3.6in]{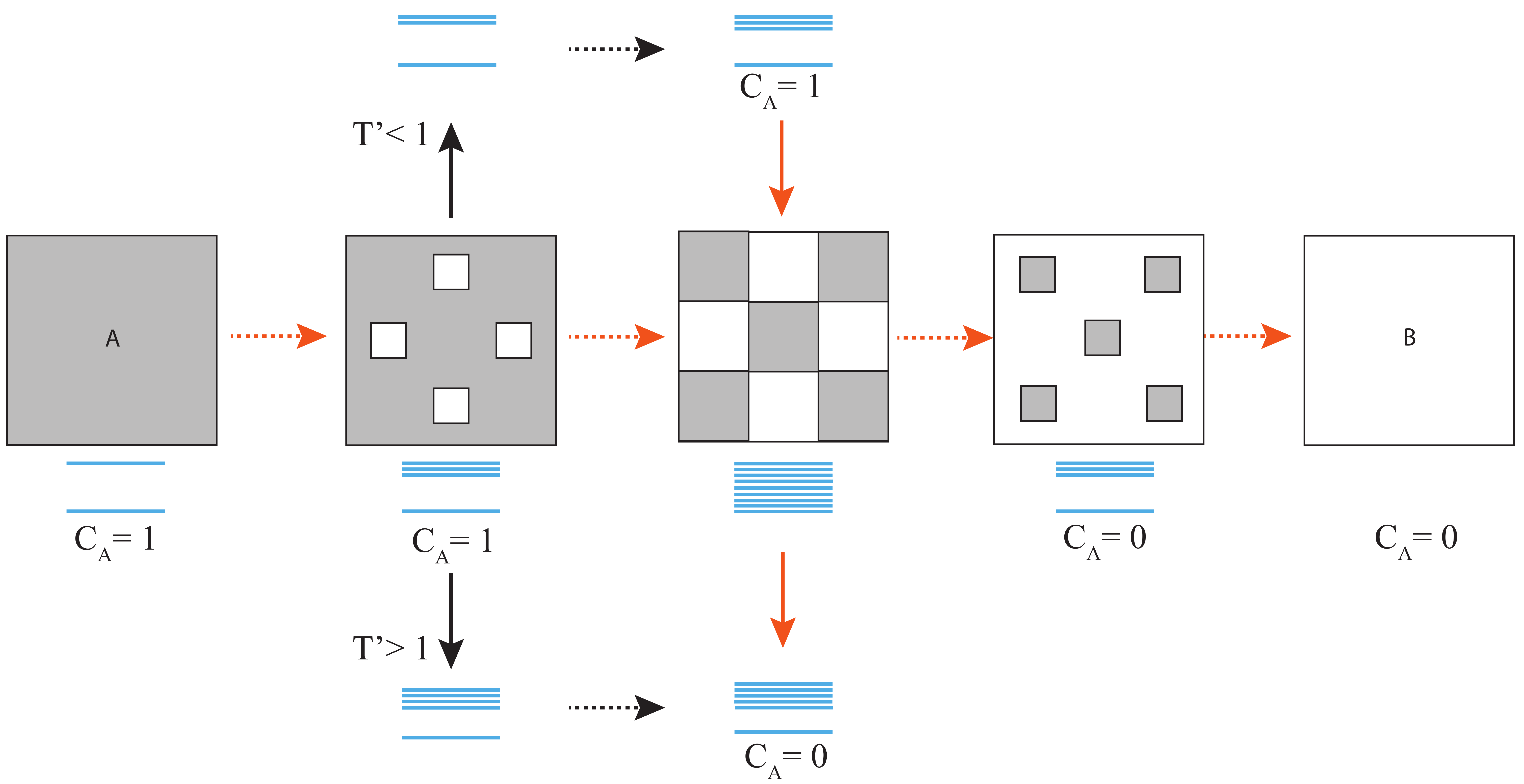}
\caption{Extensive partition of a topological state yields a topological phase transition in the bulk entanglement Hamiltonian $H_A$ of a subsystem. The horizontal and vertical sequences of orange arrows represent two ways of realizing the 
transition:  the horizontal sequence denotes geometrically tuning the partition towards quantum criticality at the symmetric point (center); the vertical sequence comes from changing the entanglement temperature ($T=1\rightarrow T'$) at an earlier stage of an asymmetric partition. Dotted arrows always indicate the tracing out procedure. The schematic bulk entanglement spectrum and topological invariant for $A$ are shown in blue at every stage of partition. $C_A=1$ denotes the topological order of the original topological ground state, and $C_A=0$ denotes topologically trivial.}
\end{figure*}

The entanglement between two parts, $A$ and $B$, of a many-body ground state $|\Psi \rangle$  is characterized by the reduced density matrix $\rho_A$.  $\rho_A$ can be formally written as the thermal density matrix of an entanglement Hamiltonian $H_A$ at temperature $T=1$: 
\beq
\rho_A = {\textrm Tr}_B |\Psi\rangle\langle\Psi| \equiv  e^{-H_A }.   \label{rho}
\eeq 
The full set of eigenvalues of $H_A$, denoted by $\{  \xi_i \} $ with $1 \leq i  \leq {\rm dim} [ {\cal H}_A ] $,  constitutes the entanglement spectrum of subsystem $A$.
These eigenvalues are directly related to the coefficients in the Schmidt decomposition of the ground state:
\beq
|\Psi\rangle = \sum_i e^{-\frac{\xi_i}{2}} |\psi_i \rangle_A \otimes |\tilde{\psi}_i \rangle_B.  \label{schmidt}
\eeq 
Among all states in subsystem $A$, we will pay attention to those ``dominant states'' that have small eigenvalues and hence large weight in the ground state $|\Psi\rangle$.   

When $A$ and $B$ are the left and right halves of a quantum Hall state, the entanglement spectrum is gapless, and its low-lying part resembles the excitation spectrum of a physical edge\cite{lihaldane, bernevig, qi, swingle, pollman, fid, yao, dubail}. 
Due to the inherent boundary-local nature\cite{read, locality}, the entanglement spectrum from the left-right partition does not directly reveal properties of the bulk. 

We introduce an extensive partition to study the bulk:  a system is divided into two subsystems $A$ and $B$ consisting of a periodic array of blocks. See Fig. 1 for examples.
The shape of each block does not matter; the defining characteristic of such extensive partitions is that the entire boundary between $A$ and $B$ 
extends throughout the bulk of the system in all directions. Because (i) quantum entanglement in a gapped system comes mostly from boundary degrees of freedom and (ii) 
the boundary between $A$ and $B$  in an extensive partition is itself extensive, the low-lying part of the corresponding entanglement spectrum contains information about the bulk. 

To demonstrate this bulk nature explicitly, we note that the entanglement entropy of an extensive partition $S=- {\rm Tr} (\rho_A \log \rho_A )$ 
scales  as the {\it total area} of the boundary:  $L^{d-1} N$, where $L$ is the linear size of each block, $d$ is the spatial dimension of the system, and $N$ is the number of blocks in the entire system. 
In the thermodynamic limit, $N \rightarrow \infty$ while $L$ is fixed. Therefore $S$ is proportional to the {\it volume} of the system: $S \sim V/L \propto V$. 
In this case, it is meaningful and instructive to draw analogies between the reduced density matrix $\rho_A$ and the thermal density matrix 
of a $d$-dimensional physical system defined on $A$ (a superlattice), and between the entanglement Hamiltonian $H_A$ and the physical Hamiltonian.


We will study a sequence of extensive partitions for a topological state. 
Depending on the actual partition, the bulk entanglement spectrum  can be either gapped or gapless.   
When the BES is gapped,  its lowest eigenvalue $\xi_0$  is separated from the rest  
by a finite amount in the thermodynamic limit.  
In this case, the corresponding product state $| \psi_0 \rangle_A \otimes | \tilde \psi_0\rangle_B$, which has the largest weight in the Schmidt decomposition, 
stands out as the most dominant component in the ground state $|\Psi\rangle$. 
Here $|\psi_0\rangle_A$ is the lowest eigenstate of the entanglement Hamiltonian of subsystem $A$, 
which is defined by $H_A | \psi_0 \rangle_A = \xi_0 | \psi_0 \rangle_A$ and hereafter referred to as entanglement ground state; likewise $|\psi_0\rangle_B$ is the entanglement ground state of subsystem $B$. 
In contrast, when the BES is gapless,  
no individual product state can be singled out as the most dominant component in $|\Psi \rangle$.    

We now derive the main result of this work:  by varying the extensive partition between two limits defined below, the bulk entanglement Hamiltonian $H_A$ of a given subsystem in a topological state undergoes a gap-closing transition, in which the entanglement ground state $|\psi_0 \rangle_A$ changes from being topologically equivalent to the nontrivial ground state $|\Psi\rangle$ to being trivial.  Throughout the following general discussion, it will be useful to keep in mind a concrete example of a topological phase with gapless edge excitations, such as a quantum Hall state.  For such phases, the low-lying states in the entanglement spectrum of subsystem $A$ arise from boundary degrees of freedom that correspond to a network of chiral edge states, which would occur if $B$ had been physically removed (Fig. 2a).  Varying the extensive partition changes how these edge modes interact with each other, and we will refer to this `edge picture' when relevant.

First consider extensive partitions in two extremely asymmetric limits: 
(i) a percolating sea of region $A$ with an array of small, isolated islands of region $B$, and (ii) a percolating sea of region $B$ with an array of small, isolated islands of region $A$. We require that  different islands are separated by  a distance  $L$ which is greater than the correlation length $\xi$ in the ground state. 
Since a given island has a finite number of degrees of freedom, the entanglement spectrum of this island, obtained by tracing out the rest of the system, 
generically has a unique ground state with a finite gap. 
Moreover, since $L \gg \xi$, different islands contribute to the entanglement between $A$ and $B$ independently. 
Therefore the BES from extensive partitions (i) and (ii) are both gapped.  In the edge picture, the edge modes are confined within each block and do not percolate (Fig. 2c). For such Chalker-Coddington network models, the spectrum is gapped\cite{cc}.

On the other hand, the entanglement ground states in partitions (i) and (ii) are vastly different. 
Since the density matrix $\rho_A$ in (ii) is defined on disconnected islands,  it factorizes into a direct product and 
thus must be topologically trivial.   In contrast,  since $\rho_A$ in (i) is defined on the percolating sea, 
it must have the same topological order as the ground state $|\Psi\rangle$:  this is manifest in the extreme limits where $B$ is the null set.  Hence, as one shrinks the size of $A$ and concurrently enlarges $B$ to interpolate from partition (i) to (ii), the entanglement ground state 
$|\psi_0\rangle_A$ must change from carrying the topological order to being a  trivial product state of islands. 
Whether topological order is present in a gapped system is a yes or no question. Therefore 
in order to accommodate this change of topology in the entanglement ground state, 
the bulk entanglement Hamiltonian  $H_A$  must close the gap somewhere in between partitions (i) and (ii).      

Where does the gap close? We now argue that 
for the vast majority of topological states (to be precisely defined below), the gap in the bulk entanglement spectrum must close at 
a symmetric partition, where $A$ and $B$ are related by symmetry such as translation or reflection (see Fig.1c or 2a).  In the edge picture, for such symmetric extensive partitions the edge modes have equal left- and right-turning amplitudes at the nodes, and these network models are critical\cite{cc}.  

To understand this claim more generally, we first make the following conjecture: if the bulk entanglement spectrum from an extensive partition is gapped, the largest-weight state 
$| \psi_0 \rangle_A \otimes | \tilde \psi_0\rangle_B$ in the Schmidt decomposition possesses the same topological order as the original ground state $| \Psi \rangle$.  
Intuitively, this conjecture is expected to hold because all other components in the Schmidt decomposition,   
$|\psi_i\rangle_A \otimes | \tilde{\psi}_i\rangle_B$ with $i \neq 0$, 
are exponentially suppressed by the entanglement gap.  
Therefore it should be possible to ``push'' the entanglement gap to infinity\cite{thomale} and 
thereby deform the original ground state (\ref{schmidt}) into the entanglement ground state of the extensive partition $|\psi_0\rangle_A \otimes | \tilde{\psi}_0\rangle_B$.  
This argument suggests that $| \psi_0 \rangle_A \otimes | \tilde \psi_0\rangle_B$ and $|\Psi \rangle$ 
are adiabatically connected and hence carry the same topological order. 

Let us apply the above conjecture to a symmetric partition of a topological state $|\Psi\rangle$, where $A$ and $B$ are symmetry-related. 
Suppose for the sake of argument that the bulk entanglement spectrum is gapped.  It then follows from our conjecture that 
$|\Psi \rangle$ is topologically equivalent to a direct product of two essentially identical states: $|\psi_0\rangle_A \otimes | \tilde{\psi}_0\rangle_B$, where   
$|\psi_0\rangle_A$ and $| \tilde{\psi}_0\rangle_B$ are related by a translation or reflection. This implies that 
$|\Psi \rangle$ must have a ``doubled'' topological order. In contrast, 
the majority of topological states carry an elementary unit of topological order that cannot be divided equally into halves.  
We thus prove by contradiction that the bulk entanglement spectrum of an ``irreducible'' topological state must be either gapless or have degenerate ground states at a symmetric partition.   

To summarize, we have argued that 
the discrete nature of topological order dictates that one subsystem in an extensive partition inherits the topological order while the other does not. 
As a consequence, as one varies the partition, phase transition(s) must occur in the bulk entanglement spectrum, which is constructed from a single ground state wavefunction.  Our conclusion is based on general principles as well as the microscopic understanding in terms of percolating edge modes. 
 This is illustrated below using the example of integer quantum Hall states, 
for which {\it rigorous} results will be derived.

{\bf Example}: We consider an integer quantum Hall state on a two-dimensional lattice\cite{haldane} (also known as the Chern insulator), defined by a generic, translationally-invariant 
tight-binding Hamiltonian $H$. 
For such free fermion systems, the entanglement Hamiltonian of a subsystem $A$ takes a quadratic form\cite{peschel}: $H_A=\sum_{r, r' \in A} {\cal H}_A(r,r')  c^\dagger_r c_{r'} $. The  
set of eigenvalues of $\cal H$ is denoted by $\{\epsilon_i\}$, which can be regarded as single-particle levels. 
The entanglement spectrum of $A$ is then obtained by  filling these levels. It then follows that the important, low-lying part of the entanglement spectrum for a fixed density of particles in $A$ comes from the vicinity of the highest occupied level. 

The  form of $H_A$ is entirely determined from the two-point correlation function of the ground state $C(r, r') = \langle \Psi | c^\dagger_r c_{r'} | \Psi \rangle$, where both sites 
$r$ and $r'$ lie within subsystem $A$.  
Using the exponential decay of $C(r,r')$ at large distance $|r-r'|$, we prove\cite{footnote} that $H_A$ is short-ranged and resembles a physical Hamiltonian, provided that the volume of $A$ does not exceed that of $B$.  In the opposite case, $H_A$ has bands at $\pm \infty$\cite{footnote}; nonetheless, the low-lying part of the spectrum still resembles the excitation spectrum of a bulk, physical Hamiltonian.  

The ground state of an integer quantum Hall system is indexed by a nonzero Chern number \cite{tknn}: $C\neq 0$. 
As a consequence, one cannot choose the phase of Bloch wavefunctions continuously over the entire Brillouin zone in reciprocal space.  
Based on this topological obstruction, we prove\cite{footnote} that the bulk entanglement spectrum from a symmetric extensive partition 
must be gapless when $C$ is an {\it odd} integer (but not necessarily so when $C$ is even). Such an 
entanglement criticality at the symmetric partition of {\it irreducible} quantum Hall states agrees with our conclusion deduced earlier from general considerations.  


\begin{figure}
\centering
\includegraphics[height=2.5in]{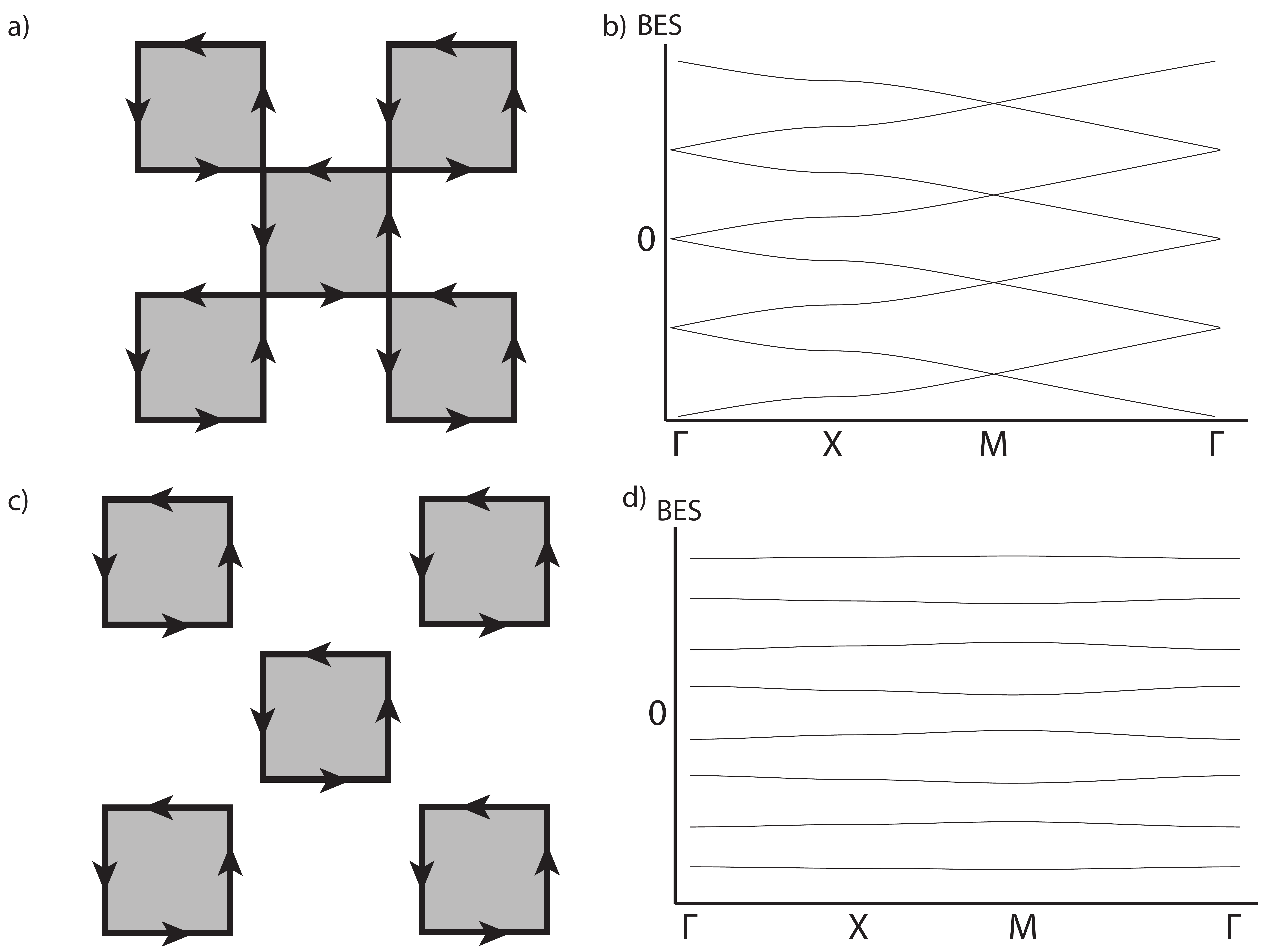}
\caption{(a) At the symmetric partition of the integer quantum Hall state, the low-lying states in the bulk entanglement spectrum percolate. (b) The corresponding single particle bulk entanglement spectrum for the symmetric partition of the integer quantum Hall state. The model Hamiltonian used here is 
$H(k) = (\cos{k_x}+\cos{k_y}-\mu)\sigma_z + \sin{k_x} \sigma_x + \sin{k_y} \sigma_y$.
 Each block of the partition used is 5 by 5 sites. Note the massless Dirac dispersion at $\Gamma$. (c) An asymmetric partition in which the low-lying states do not percolate.  The Hilbert space is the same as that of a), but the lattice constant is different. The resulting BES is gapped (d).}
\end{figure}

To gain further insight into the bulk entanglement spectrum, 
we choose a particular model of a $C=1$ integer quantum Hall state on a square lattice, and study extensive partitions in which $A$ and $B$ are two sets of blocks as shown in Fig.2a. The single-particle levels $\{\epsilon_j\}$ that define the 
BES of A are plotted as a function of crystal momentum in the 2D Brillouin zone $0<k_x, k_y < 2\pi/L$ in Fig.2. 
We find numerically  that when $A$ and $B$ are asymmetric, the BES is gapped and the entanglement ground state of 
the (non-)percolating subsystem carries Chern number $C=1$ (0). 
At the symmetric partition, the bulk entanglement Hamiltonian exhibits a two-dimensional massless Dirac fermion spectrum (Fig.2b), 
which precisely coincides with a physical Hamiltonian tuned to the quantum Hall transition point at a fixed density and in the absence of disorder.

The correspondence between entanglement criticality and quantum Hall plateau transition is remarkable and can be understood as percolation of chiral edge states, as noted earlier.  In this microscopic sense, the phase transition in the bulk entanglement spectrum mirrors the percolation of edge states in an actual quantum Hall  transition, both of which are described by the Chalker-Coddington network model and exhibit a massless Dirac fermion spectrum at criticality\cite{cc}. 

The entanglement phase transition can be achieved by tuning the partition in multiple ways. 
So far, we have changed the size of blocks to vary the degree of asymmetry between $A$ and $B$. In this procedure, 
$A$ and $B$ become switched across the symmetric point, which leaves the entanglement spectrum invariant (but not the entanglement ground state).  
This leads to an interesting {\it duality} relating the two sides of the entanglement critical point. 

On the other hand, as the size of blocks in subsystem $A$ changes,  the Hilbert space of the bulk entanglement Hamiltonian $H_A$ changes accordingly, 
which is rather different from typical quantum phase transitions.  
An alternative way of tuning an extensive partition without changing the Hilbert space dimension of $A$ is to   
start from an array of disconnected blocks, keep the size of blocks fixed, and decrease the distance between blocks in $A$ up to the symmetric point where adjacent blocks touch at corners. Beyond the symmetric point, different blocks touch on edges to make $A$ 
a percolating sea.    

Last but not least, we introduce a third procedure to achieve entanglement phase transition on a {\it fixed} subsystem $A_s$ of a symmetric partition. This is achieved by explicitly constructing a {\it continuous} family of bulk entanglement Hamiltonians $H_{A_s}(T)$ at different ``entanglement temperature'' $0<T<\infty$. 
$H_{A_s}(T)$ is defined as follows: 
\beq
 e^{-H_{A_s}(T)} &\equiv& \frac{1}{Z} \textrm{Tr}_{B_s -B } [ e^{-  H_{A}/T} ], \nonumber \\
\text{where}  \;  \; \;  e^{-H_{A}} &=& \textrm{Tr}_{B} |\Psi \rangle \langle \Psi |, \nonumber \\ 
  Z &=& \textrm{Tr}(e^{- H_A/T}).
\eeq 
In words, we first trace out a subregion of $B_s$ (the complement of $A_s$), denoted by $B$. This yields an entanglement Hamiltonian $H_A$ for the complementary subsystem $A$. Next, we change the entanglement temperature $T$ to construct a new 
density matrix $\rho_A (T)   \equiv Z^{-1} e^{- H_A/T}$, and then continue tracing out remaining sites in $B_s$. This procedure yields a $T$-dependent reduced density matrix for $A_s$, from which $H_{A_s}(T)$ is defined.  

Our protocol is inspired by the following intuition:  raising the entanglement temperature achieves a similar effect as tracing out sites in that they both reduce the amount of information in a subsystem.  
$H_{A_s}(T=1)$ reduces to  $H_{A_s}$ defined previously by tracing out $B_s$ directly in the symmetric partition, and hence is right {\it at} the critical point. 
On the other hand, raising the entanglement temperature above $T=1$ at an intermediate partition is equivalent to tracing out more than $B_s$ in the original scheme, which generates a trivial state with $C_{A_s}(T>1)=0$.  
In contrast, lowering the entanglement temperature ``purifies'' the topological order: in the extreme limit $T=0$, we are simply taking the ground state of $H_A$ at the asymmetric partition (with $C_A=C=1$) as a new starting point and then tracing out a {\it minority} region to reach the symmetric partition. As a result, the ground state of $H_{A_s}(T<1)$ inherits the nonzero topological invariant of $H_A$.  
We have verified this numerically:  $H_{A_s}(T)$ is gapped for $T \neq 1$, and the entanglement ground state 
has Chern number $C_{A_s}=1$ for $T<1$ and $C_{A_s} = 0$ for $T>1$. Therefore $H_{A_s}(T)$ 
undergoes the topological quantum phase transition as a function of the entanglement temperature $T$, with critical point at $T=1$.

{\bf Discussion}:  While our general argument proves the existence of 
a phase transition in the bulk entanglement spectrum induced by varying the partition, the question remains 
whether this ``entanglement phase transition'' exhibits the same critical properties as a phase transition induced by tuning physical parameter in a physical Hamiltonian.   
This is indeed the case for integer quantum Hall states, as shown earlier. What about topological phases in general?    
 
Before discussing this issue, it should be noted that the trivial state that appears in the bulk entanglement spectrum of an asymmetric partition is a direct product of blocks. Thus the entanglement phase transition should be compared with an actual phase transition into such a product state, which can often be realized by imposing an external periodic potential.       
We expect that the entanglement phase transition obtained from extensive partition exhibits the same critical properties 
as actual transitions, if the latter are continuous and fall into a single universality class.  
Alternatively, both entanglement and actual phase transitions into block-product states can be first order or intervened by an extended critical phase. 

Finally, we end by discussing {\it why} a topologically nontrivial ground state is capable of ``knowing'' its own phase transition. 
We believe this is due to the peculiar nature of topological phases and topological phase transitions. Without any local order parameter, topological phases  
are distinct from trivial states only because of nontrivial patterns of entanglement\cite{wenxie}. For this sole reason, a transition is needed to go from one phase to the other. 
Therefore, it appears to us that a topological phase transition can be generically regarded as percolation of entanglement, which is a generalization of edge state percolation at the quantum Hall transition.  In this sense, our work generalizes readily to other systems with gapless edge excitations in any symmetry class and dimension\cite{periodictable}.  On the other hand, topological phases such as $Z_2$ spin liquids do not necessarily have gapless edge states. It remains an  interesting and open question whether or not phase transitions in these systems can be understood from the perspective of entanglement percolation.

{\it Acknowledgement:} 
We thank Xiao-Liang Qi for an important and stimulating discussion, as well as Karen Michaeli, Patrick Lee, Senthil Todadri, Jan Zaanen, Xiao-Gang Wen, Guifre Vidal, Frank Wilczek, Andreas Ludwig, Brian Swingle, Maissam Barkeshli and Adam Nahum for interesting comments and thought-provoking questions. We also thank an anonymous referee for valuable suggestions. 
LF is partly supported by the DOE Office of Basic Energy Sciences, Division of Materials Sciences and Engineering under award DE-SC0010526. LF thanks the Institute of Advanced Study at the Hong
Kong University of Science and Technology and University of Tokyo, where part of this work was conceived. 
TH is supported by NSF Graduate Research Fellowship No. 0645960 and thanks the Perimeter Institute, where part of this work was completed.

\section{Supplementary Material}

We first review the derivation of the entanglement Hamiltonian for free fermion systems, before applying it to extensive partitions.  Then, we prove that an extensive partition of an odd Chern number free fermion ground state yields a gapless bulk entanglement spectrum.  Finally, we establish the locality of the entanglement Hamiltonian for a certain class of extensive partitions.

\subsection{Entanglement Hamiltonian from Extensive Partitions of Free Fermion Ground States}
As shown by Peschel \cite{peschel}, for ground states which are Slater determinants, the reduced density matrix for a subregion $A$ is simply related to the two point correlation functions on $A$.  This is because the reduced density matrix is defined to reproduce all observables on $A$, and Wick's theorem allows all higher-order correlation functions to be derived from two-point functions.  Hence, the reduced density matrix can be written as
\beq
\rho_A &=& \frac{1}{Z}e^{-{\cal H}} \\
{\cal H} &=& \sum_{i,j} H_{ij} c_i^{\dagger} c_j,
\eeq
where $i,j$ label sites in the subsystem $A$ of interest, and $Z=tr{e^{-{\cal H}}}$.

Let the eigenvalues and eigenvectors of $H$ be $\epsilon_k, \phi_k(i)$.  Then
\beq
H_{ij} = \sum_k \phi_k (i) \phi_k^* (j)\epsilon_k.
\eeq
The correlation matrix $C_{A,ij}\equiv Tr(\rho_A c_i^{\dagger} c_j)$ is
\beq
C_{A,ij} = \sum_k \phi_k^* (i) \phi_k (j) \frac{1}{e^{\epsilon_k}+1}
\eeq

We emphasize that $i,j$ are restricted to region $A$.

Thus 
\beq
H^T = \ln \frac{1-C_A}{C_A}. \label{hc}
\eeq

\begin{figure*}
\centering
\includegraphics[height=1.5in]{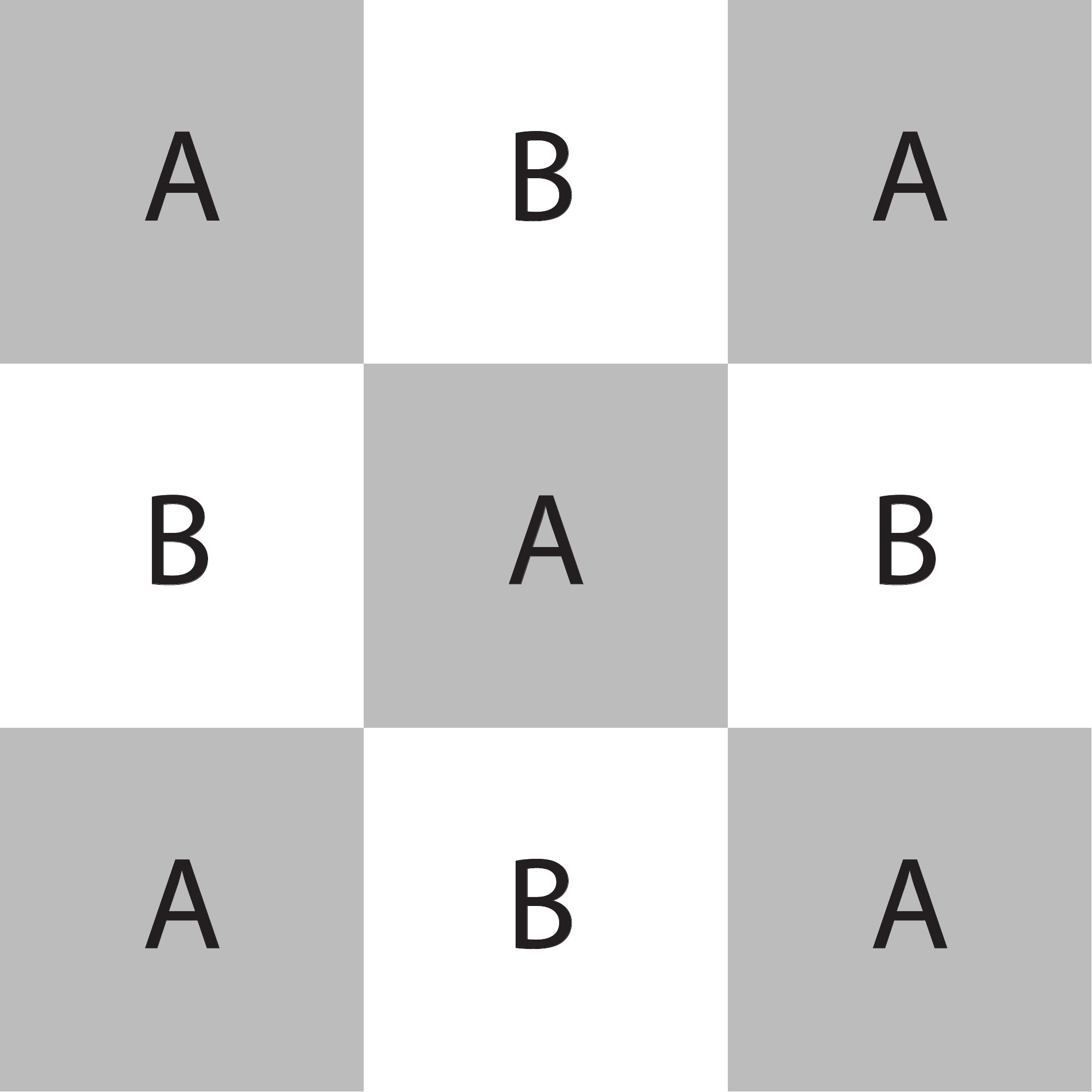}
\caption{The partition used in computing the entanglement Hamiltonian.  Each patch is an $L\times L$ block.}
\end{figure*}

We now derive the entanglement Hamiltonian from tracing out every other square block of sites from a free fermion ground state on an {\it infinite} square lattice (see Fig.1).  Begin with a ground state that is invariant under translations by one lattice spacing, and consider tracing out every other $L$ by $L$ block of sites.  We emphasize that subsystem $A$ still has translation symmetry along two directions, with new lattice vectors $(\pm L,L)$. Define the unit cell of the full system to be an $A$ block (an $L$ by $L$ square).  Then the corresponding Brillouin zone is a $\frac{2\pi}{L}$ by $\frac{2\pi}{L}$ square.  The full correlation matrix is 
\beq
C_{xx',\sigma \sigma'} &=& <c_{x\sigma}^{\da} c_{x'\sigma'}> \\
&=& \frac{1}{V} \sum_{k\in BZ} <c^{\da}_{k\sigma} c_{k\sigma'}> e^{-ik(x-x')}, 
\eeq
where $x, x'$ label unit cells, and $\sigma, \sigma'$ label orbital degrees of freedom, including basis sites within each unit cell.  

Let $\{|n_{k'}\rangle\}$ denote the set of occupied bands at momentum $k'$ in the Brillouin zone.  Then the ground state can be written as
\beq
|\psi\rangle = \prod_{k',n_{k'}} \sum_{\sigma} \langle n_{k'}|\sigma\rangle c^{\dagger}_{k' \sigma} |0\rangle.
\eeq

Then 
\beq
<\psi|c^{\da}_{k\sigma} c_{k\sigma'}|\psi> = \sum_{n_k, n'_k} \langle \sigma |n_k\rangle\langle n'_k|\sigma'\rangle.
\eeq

Hence,
\beq
C_{xx',\sigma\sigma'} = \frac{1}{V} \sum_{k\in BZ} \sum_{n_k,n'_k} \langle \sigma |n_k\rangle\langle n'_k|\sigma'\rangle e^{-ik(x-x')}.
\eeq
Suppressing the orbital indices,
\beq
C_{xx'} = \frac{1}{V} \sum_{k\in BZ} |\psi_k\rangle\langle \psi_k| e^{-ik(x-x')}
\eeq
where $|\psi_k\rangle \equiv \prod_{n_{k}} \psi^{\dagger}_{n_k} |0\rangle \equiv \prod_{n_k} \sum_{\sigma} \langle n_{k}|\sigma\rangle c^{\dagger}_{k \sigma} |0\rangle$.

Now we compute the correlation matrix $C_A$ restricted to subsystem $A$.  {\it We emphasize that the subsystem $A$ has translational symmetry along both directions, with new lattice vectors $(\pm L,L)$}.  Hence, the eigenvalues and eigenvectors of $C$ are still labeled by momentum $\tk$, now restricted to a reduced Brillouin zone (see $R$ in Fig.2) because $x,x'$ are restricted to sublattice $A$.  Note that $C_A$ acting on a plane wave $|\tk\rangle \equiv e^{-i\tk x}$ yields

\beq
C_{A}|\tk\rangle=\frac{1}{V} \sum_{x'} \sum_{k\in BZ} |\psi_k \rangle\langle \psi_k| e^{-ikx}e^{ix'(k-\tk)} \\
= \frac{1}{2} (|\psi_\tk\rangle\langle \psi_\tk|+|\psi_{\tk+Q}\rangle\langle \psi_{\tk+Q}|) |\tk\rangle
\eeq

Hence, the Fourier transformed restricted correlation matrix is
\beq
C_{A, \tk} 
= \frac{1}{2} (|\psi_\tk\rangle\langle \psi_\tk|+|\psi_{\tk+Q}\rangle\langle \psi_{\tk+Q}|) \label{cproj}
\eeq
where $Q \equiv (\pi/L,\pi/L)$.  We remind the reader that $C_{A,\tk}$ has matrix elements $C_{A,\tk,\sigma\sigma'}$ obtained by sandwiching the above expression between $\langle \sigma|$ and $|\sigma'\rangle$.

\subsection{Gapless Entanglement Hamiltonians for Symmetric Partitions of Odd Chern Number Insulators}
If the free fermion ground state has odd Chern number, then the symmetric partition yields a gapless entanglement Hamiltonian, as we now show.  We first establish that 
when $A$ and $B$ are related by translational symmetry, the entanglement spectrum has particle-hole symmetry (see \cite{turner} for a discussion in the case of inversion symmetry).  One can understand this as follows: $C$ has single-particle eigenmodes $\{m^A_j\}$ corresponding to eigenvalues $\{p_j\}$ between 0 and 1.  Then in the decomposition  of the $N$-particle ground state $|\Psi\rangle$
\beq
|\Psi\rangle = \sum_i e^{-\frac{\xi_i}{2}} |\psi_i \rangle_A \otimes |\tilde{\psi}_i \rangle_B.  \label{schmidt}
\eeq
each state $|\psi_i \rangle_A$ is a choice of which $n$ modes of $\{m^A_j\}$ to occupy (with $n<N$).  $p_j$ represents the weight of the corresponding mode to appear in the decomposition of the ground state \cite{klich}:
\beq
|\Psi\rangle = \prod_j \big(\sqrt{p_j} \psi^{\dagger}(m^A_j)+\sqrt{1-p_j}\psi^{\dagger}(m^B_j)\big)|0\rangle .  \label{schmidt}
\eeq

Applying the translation operator relating $A$ to $B$ yields 
\beq
|\Psi\rangle = \prod_j \big(\sqrt{p_j} \psi^{\dagger}(\tilde{m}^B_j)+\sqrt{1-p_j}\psi^{\dagger}(\tilde{m}^A_j)\big)|0\rangle . 
\eeq

Thus, every mode with eigenvalue $p_j$ has a corresponding mode with eigenvalue $1-p_j$. This establishes particle-hole symmetry of $C$ and hence $H$.

Therefore, to show that $H_A$ is gapless, it suffices to show that $C$ has eigenvalue $1/2$.  To do this, we show that there must be a momentum $k_0$ in $R$ at which $|\psi_{k_0}\rangle$ and $|\psi_{k_0+Q}\rangle$ are orthogonal, in which case (\ref{cproj}) ensures an eigenvalue $1/2$.  

Assume for the sake of contradiction that this is not true.  Then we can smoothly choose a gauge for an open patch containing $R$ and then smoothly choose a gauge for an open patch containing $S$ such that 
\beq
\langle \psi^R_k|\psi^S_{k+Q}\rangle>0 \label{condition}
\eeq
On the diamond boundary $\partial R$ between $R$ and $S$, the two sets of wavefunctions differ by a phase:
\beq
|\psi^S_k\rangle = e^{i\theta_k} |\psi^R_k\rangle, k\in \partial R
\eeq

But our choice of gauge in equation $\ref{condition}$ implies that 
\beq
e^{i\theta_k}=\frac{\langle \psi^S_k | \psi^S_{k+Q}\rangle}{|\langle \psi^S_k | \psi^S_{k+Q}\rangle|}. \label{winding}
\eeq

The Chern number is the winding number of $\theta_k$ around $\partial R$ \cite{kohmoto}.  However, the winding of the right side of \ref{winding} can be divided into two closed paths $abc$ and $cda$ (because $a,c$ and $b,d$ are the same points).  Each loop yields the same winding number because $\langle \psi^S_k | \psi^S_{k+Q}\rangle=\langle \psi^S_k | \psi^S_{k+Q}\rangle^*$, while the paths are traversed in opposite directions.  This implies that the Chern number is even.  By contradiction, there must be a momentum $k_0$ in $R$ at which $|\psi_{k_0}\rangle$ and $|\psi_{k_0+Q}\rangle$ are orthogonal, and hence $C$ and $H$ are gapless.

\begin{figure*}
\centering
\includegraphics[height=1.5in]{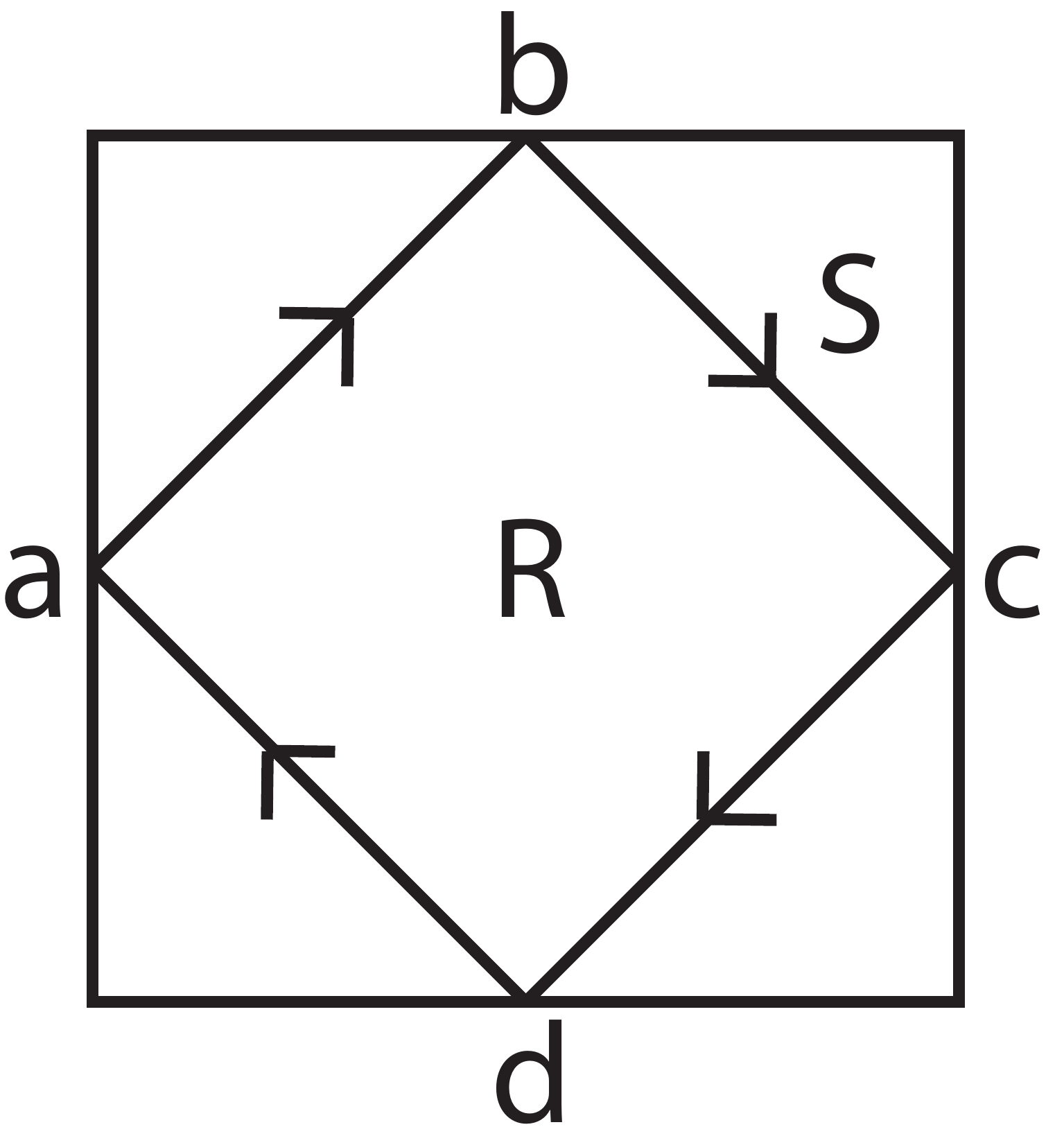}
\caption{(big square) Folded Brillouin zone for the coarse-grained square lattice (Fig.1). $R$ (inner diamond) is the reduced Brillouin zone for the $A$ sub lattice. $S$ is its complementary region.}
\end{figure*}

We have checked that in a ground state with Chern number 2, the BES for some symmetric partition is gapped.

\subsection{Properties of Entanglement Hamiltonian from Extensive Partitions}
We now discuss to what extent the entanglement Hamiltonian for a gapped free fermion ground state resembles a physical Hamiltonian, for the extensive checkerboard-type partitions considered in our work. 
Note that the correlation matrix $C$ is local ($C_{ij} \propto e^{-\alpha |i-j|}$) because it originates from the ground state of a gapped system.  
$C$ and $H_A$ have the same eigenstates but different eigenvalues (\ref{hc}).  
First assume that $C$ does not have eigenvalues 0 or 1, i.e. when there are no states in $A$ that are definitely or definitely not occupied.  The fact that $C$ decays exponentially in real space implies that its Fourier transform $C(k)$ is an analytic function of $k$ in a strip in the complex $k$ plane infinite in the real direction and of finite width $w$ in the imaginary direction \cite{analytic}.  Since we are assuming $C$ does not have eigenvalues 0 or 1, $H_A(k)= \ln \frac{1-C(k)}{C(k)}$ is also an analytic function of $k$ in the same strip.  This implies \cite{analytic} that $H_A$ also decays exponentially in real space. Therefore $H_A$ genuinely resembles a bulk Hamiltonian defined on subsystem $A$.  

The above assumption that $C$ does not have eigenvalues 0 or 1 is generically true when subsystem $B$ can accommodate all the particles of the ground state. 
On the other hand, when $A$ is larger than $B$, there will be states in $A$ that are always occupied in the Schmidt decomposition of the physical ground state, since $B$ cannot contain the original number of particles in the ground state.  The eigenvalues 0 and 1 translate to $\pm \infty$ eigenvalues of $H_A$.  
However, the low-lying part of the entanglement spectrum  arises from excitations near the highest occupied band, and thus still resembles the excitation spectrum of a local Hamiltonian.

\end{document}